# A refractive index sensor based on the leaky radiation of a microfiber

F. Gao, H. Liu,[*] C. Sheng, C. Zhu and S. N. Zhu

*National Laboratory of Solid State Microstructures, School of Physics, Nanjing University, Nanjing 210093,China*
[*]*liuhui@nju.edu.cn*
*http://dsl.nju.edu.cn/dslweb/HuiPage/home.htm*

**Abstract:** In this work we present a refractive index sensor based on the leaky radiation of a microfiber. The 5.3um diameter microfiber is fabricated by drawing a commercial optical fiber. When the microfiber is immersed into a liquid with larger refractive index than the effective index of fiber mode, the light will leak out through the leaky radiation process. The variation of refractive index of liquid can be monitored by measuring radiation angle of light. The refractive index sensitivity can be over 400 degree/RIU in theory. In the experiment, the variation value 0.001 of refractive index of liquid around this microfiber can be detected through this technique. This work provides a simple and sensitive method for refractive index sensing application.

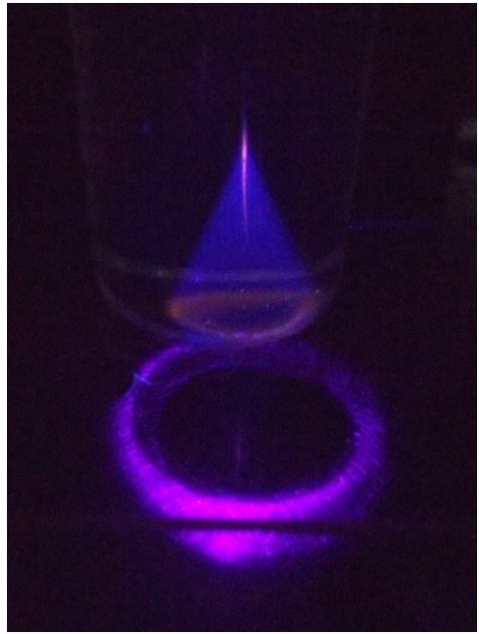

## 1. Introduction

Refractive index (RI) sensors have many applications in the field of biological and chemical. In recent years, optical fiber sensors for RI sensing have got many attentions. They have many important advantages such as small size, portable, and high sensitivity. Many kinds of fiber sensor have been previously proposed such as surface-plasmon-resonance-based fiber-optic sensor [1-3], interferometer-based fiber-optic sensors [4-7], intensity-based fiber-optic sensors [8,9], spectrally based fiber-optic sensor [10-16] and photonic crystal fiber sensors [17-19]. All these pioneering work brings us a lot of useful fiber sensor applications.

For most of applications of optical microfiber, the RI of environment medium is smaller than the effective RI of fiber modes. Then the light can be well confined inside the microfiber with very small radiation loss. But when the RI of the environment medium is larger than effective RI of the microfiber, the light will leak out and the microfiber cannot transport light in an usual way. Such a phenomenon is unuseful and should be avoided in most applications. However, in some special circumstances, leaky radiation effect can also have some interesting applications. In this work, the leaky radiation of microfiber is used to detect the RI of environment medium.

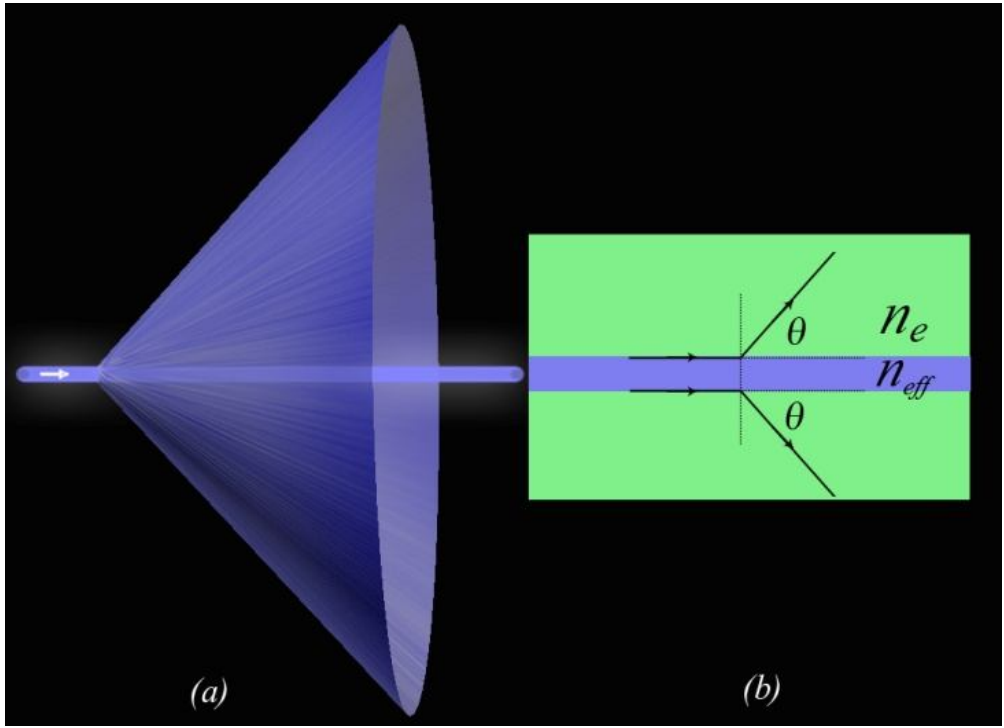

Fig. 1. The microfiber leaky radiation (a) the leaky radiation. pattern. (b) the leaky radiation angle.

The leaky radiation of microfiber is shown in Fig. 1(a). Here, the RI of environmental medium are defined as $n_e$. The effective RI of the optical fiber mode is defined as $n_{eff}$. If the RI of environment medium is larger than that of fiber mode $n_e > n_{eff}$, the light inside microfiber will leak into the environment. Based on Snell's refraction law, the leaky radiation angle from microfiber can be expressed as

$$\cos\theta = \frac{n_{eff}}{n_e} \qquad (1)$$

From Eq. (1), we can see that, if the effective RI of microfiber $n_{eff}$ is known, the radiation angle $\theta$ is only determined by the environmental RI. This means that if microfiber is immersed in environmental medium with different RI, the radiation angle will be different. Then based on this effect, the RI of environmental medium can be determined through measuring the leaky radiation angle of microfiber. Such a leaky radiation provides a simple way to detect the RI of unknown medium. Here, for the microfiber, its senstivity can be deduced from Eq. (1) as

$$\eta = \frac{d\theta}{dn_e} = \frac{n_{eff}}{n_e} \frac{1}{\sqrt{n_e^2 - n_{eff}^2}} \qquad (2)$$

According to Eq. (1) and Eq. (2), the radiation angle and sensitivity will vary with RI and these corresponding dependence relationships are given in Fig. 2. Here, $n_{eff} = 1.4624$, and $n_e$ is increased from 1.4624 to 1.7650. It can be seen that $\theta$ will go up with increasing $n_e$ while $\eta$ will go down with increasing . It clearly clarifies that $\eta$ will be increased to a very large value when $n_e$ approaching $n_{eff}$. It means that the microfiber has the better sensing performance when the environment RI is closer to RI of fiber.

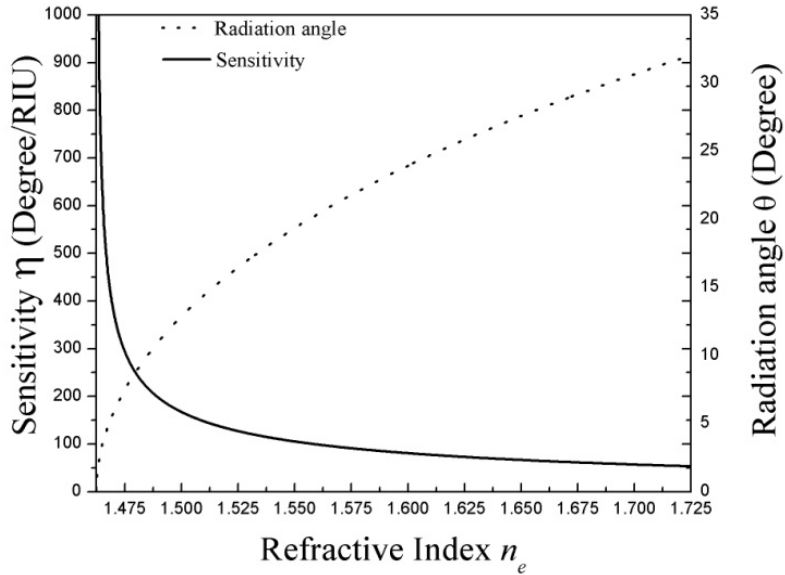

Fig. 2. The dependence of radiation angle (solid line) and sensitivity (dotted line) on RI $n_e$ (with $n_{eff} = 1.4624$).

## 2. Simulation

For purpose of proving the availability of the above model, Lumerical FDTD Solutions software is used to simulate the leaky radiation of microfiber. In the simulation model, the middle core is a 5.3um diameter microfiber and one fiber mode is excited with its effective RI $n_{eff} = 1.4624$. It is surrounded by a media with RI $n_e = 1.5000$. The simulation result is given in Fig. 3. A 405nm wavelength laser (denoted by arrows) is coupled into the microfiber and propagates from left to right. The simulation result shows that the light inside the fiber radiates out with an angle $\theta = 12.885°$. The value agrees well with the result given by Eq. (1). In other simulations with different environmental RI value $n_e$, the radiation angle $\theta$ always satisfies the relationship given in Eq. (1). Then our simulation agree well with the above leaky radiation model.

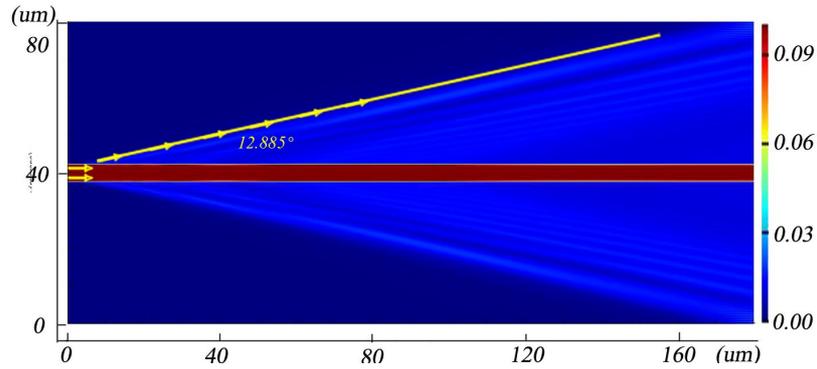

Fig. 3. Simulated leaky radiation of microfiber( $|E|^2$ profile). The radiation angle is $\theta = 12.885°$ (with $n_e = 1.5000$ )

### 3. Experimental preparations

*3.1 Fabrication of a microfiber*

The microfiber is fabricated through drawing a commercial fiber by two motorized translation stages in hydrogen flame. Before drawing the fiber, the polymer coating layer of optical fiber is stripped away. In the drawing process, the diameter of microfiber can be precisely changed by controlling the elongated length. Fig. 4 gives the picture of one fabricated microfiber through this method, which is taken with Focused ion beam (FIB). It can be seen that the 5.3um diameter microfiber surface is quite smooth and its thickness is very uniform. When laser is input into this microfiber through a fiber coupler, it will propagate along this fiber. As the polymer coating layer is removed, the light field on microfiber surface can contact with the outside environment medium directly. As a result, the light propagation will be affected by environmental medium. When the environmental index is larger than the fiber, light cannot be confined inside the fiber and leaks out. Then the leaky radiation can be observed.

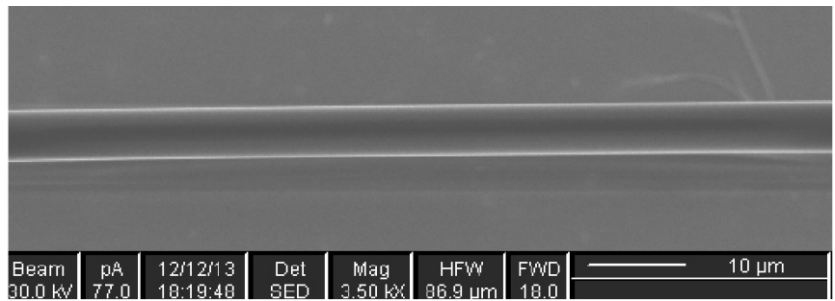

Fig. 4. FIB picture of a fabricated microfiber.

*3.2 PDMS-toluene solutions*

The PDMS(polydimethylsiloxane)-toluene mixed solution is used as environmental medium. For the pure PDMS and toluene, their RI are 1.4220 and 1.5190 at 405nm respectively. The RI of PDMS-toluene mixture is determined by the volume ratio of PDMS and toluene. Then we can change the RI of solution through changing the volume ratio of the mixture.

*3.3 Experimental setup*

In order to measure the radiation angle, an experiment setup is established as shown in Fig. 5. The solution is sandwiched between a glass substrate and wafer. A microfiber is inserted into the solution and a laser is coupled into the microfiber. As the middle liquid layer is very thin (almost equals the diameter of fiber), the radiation pattern can be seen as the projection of 3D light cone on a 2D plane. In the process, the radiation light from fiber will be scattered by the surface of wafer. Through taking the picture of the scattered signal, the leaky radiation pattern of microfiber can be obtained.

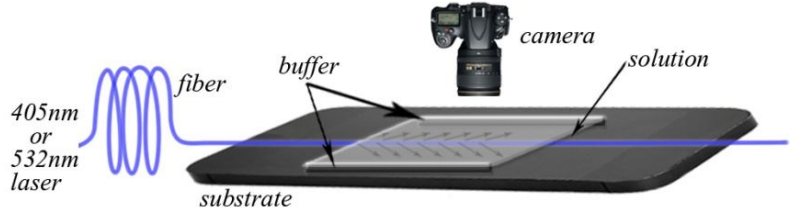

Fig. 5. Experiment setup of the leaky radiation

## 4. Experimental results and discussion

### 4.1 Experiment at 405nm wavelength

In the first experiment, 405nm laser is used to produce radiation. Eleven PDMS-toluene solutions were prepared, with different volume ratio PDMS:toluene as 1:8 , 1:6 , 1:5, 1:4 , 1:3 , 1:2.5, 1:2 , 1:1.67 , 1:1.33 , 1:1 , 1:0.75. As is anticipated in the above model and simulation, a bright light cone formed by the leaky radiation from the microfiber is observed. Fig. 6(a) shows the radiation cone (Media 1). Fig. 6(b-l) gives the pictures of radiation patterns in these different solutions taken with a CCD camera. In this 2D picture, we obtain a light fork which is the cross section of light cone on the plane. The fork angle should equal two times of the leaky radiation angle $\theta$. According to our simulations, there are many modes which can exist in the microfiber. However, in our experiment, through carefully tuning the fiber coupler, only several lower order modes are excited. Most of higher order modes are not excited in the process. The radiation cone in Fig. 6 is not a line in principle. It is composed several radiation lines for the different modes. In the process, we have to only consider the radiation line at the outmost boundary of the radiation cone. It corresponds to the highest order mode excited in our experiment. For other radiation lines immersed in the radiation cone, it is hard to determine their angles. In Fig. 6, although the two wings of light fork are broad, but the outmost boundary is quite sharp. Therefore, we can determine the radiation angle of the corresponding fiber mode through measuring the angle of sharp boundary with quite good accuracy. Five pictures were taken each time which gave a measurement for averaging. Table 1 gives the average of the measured radiation angles for the eleven solutions. It can be seen that, for the larger volume ratio of PDMS, the smaller angle is obtained correspondingly.

According to Eq. (1), the environmental RI can be calculated from

$$n_e = \frac{n_{eff}}{\cos\theta} \qquad (3)$$

Here, the RI of microfiber can be obtained as $n_{eff} = 1.4624$ from measurement of standard solution with known RI. Then the sample solutions' RI can be calculated from Eq. (3). The calculation results are provided in Table 1 and denoted as dots in Fig. 7. It shows that the smaller $n_e$ is obtained for the larger PDMS-toluene volume ratio. In this method, the measurement accuracy could reach 0.001. This high accuracy shows this method is quite sensitive compared with many other sensing techniques. In the same time, the solutions' RI is also measured by Abbe refractometer. The data from two methods are compared in Table 1 and Fig. 7. Their absolute difference value $|\Delta n|$ is also given in the bottom row of Table 1. It can be seen that two group data are quite close to each other. The maximum difference between two data is 0.0013 at the volume ration 1:8 and the minimum difference is 0.0001 at the 1:0.75. The measurement error is not larger than 0.0013. On the other hand, according to Table 1, better sensitivity is obtained for smaller environment RI. In Table 1, the overall trend shows that $|\Delta n|$ at smaller RI are less than those values at the larger RI.

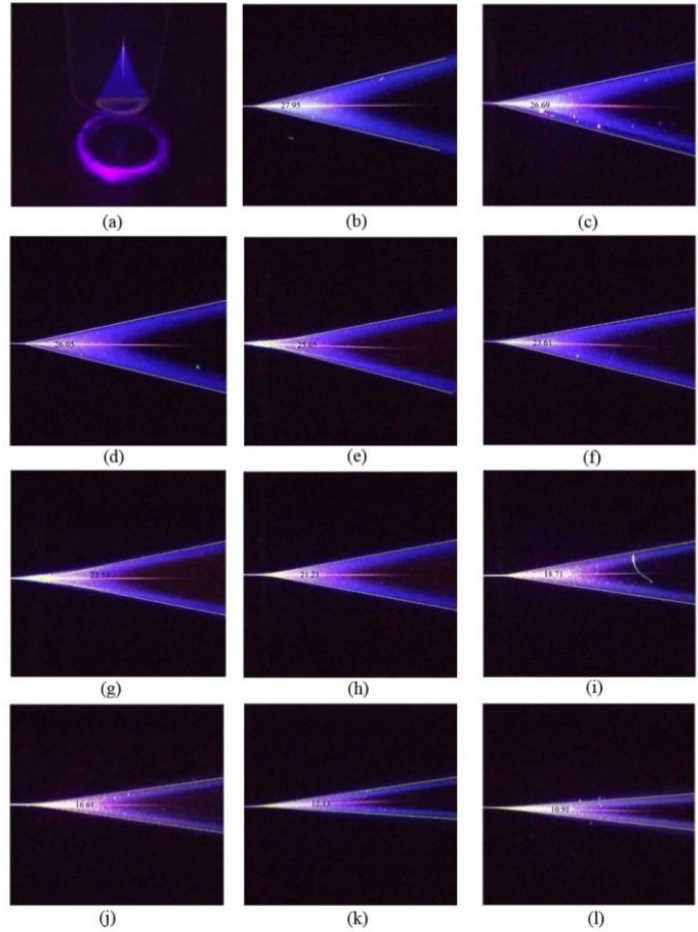

Fig. 6.(a) 405nm leaky radiation cone in the bottle (see Media 1).(b-l) 405nm leaky radiation pattern in eleven different PDMS-toluene solutions with volume ratio 1:8 , 1:6 , 1:5, 1:4 , 1:3 , 1:2.5 , 1:2 , 1:1.67 , 1:1.33 , 1:1 , 1:0.75

**Table 1. RI measured from leaky radiation and Abbe refractometer at 405nm wavelength**

| Ratio | 1:8 | 1:6 | 1:5 | 1:4 | 1:3 | 1:2.5 | 1:2 | 1:1.67 | 1:1.33 | 1:1 | 1:0.75 |
|---|---|---|---|---|---|---|---|---|---|---|---|
| M.A. | 13.990° | 13.425° | 13.015° | 12.835° | 11.765° | 11.330° | 10.635° | 9.380° | 8.315° | 6.070° | 5.480° |
| RI(L) | 1.5071 | 1.5035 | 1.5010 | 1.4999 | 1.4938 | 1.4915 | 1.4880 | 1.4822 | 1.4779 | 1.4707 | 1.4691 |
| RI(A) | 1.5084 | 1.5028 | 1.5012 | 1.4989 | 1.4940 | 1.4919 | 1.4868 | 1.4814 | 1.4782 | 1.4710 | 1.4692 |
| $\eta$ (deg/RI) | 150.27 | 161.10 | 164.60 | 170.03 | 183.49 | 190.24 | 210.08 | 239.16 | 262.97 | 358.63 | 403.92 |
| $|\Delta n|$ | 0.0013 | 0.0007 | 0.0002 | 0.0010 | 0.0002 | 0.0004 | 0.0012 | 0.0008 | 0.0003 | 0.0003 | 0.0001 |

M.A. stands for measured radiation angle in the experiments, RI(L) stands for the sample RI measured through leaky radiation, RI(A) stands for the sample RI measured through Abbe refractometer.

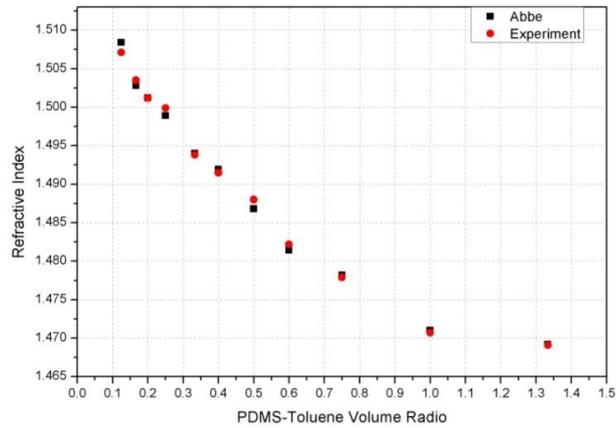

Fig. 7. Measured RI at 405nm wavelength by leaky radiation (red circle dots) and Abbe refractometer (black square dots) for different PDMS-toluene volume radio solutions.

*4.2 Experiment at 532nm wavelength*

The RI of most medium has dispersion properties. At different wavelength, their RI will be different. Then a good sensor should be able to measure RI for different wavelength. In the above experiment, 405nm laser is used to do the sensing experiment. Actually, this method is not limited to this special wavelength. In another experiment, 532nm laser is used to detect the change of RI with the same microfiber. Seven PDMS-toluene solutions with different volume ratio are used in the measurement. 532nm laser leaky radiation is shown in Fig. 8. The measured RI of the solutions by leaky radiation and Abbe refractometer are also compared in Fig. 9. Through the comparison, small measuring error and quite good sensitivity can be obtained for 532nm laser as well. In principle, this method can be used to measure the RI of medium at many other wavelengths.

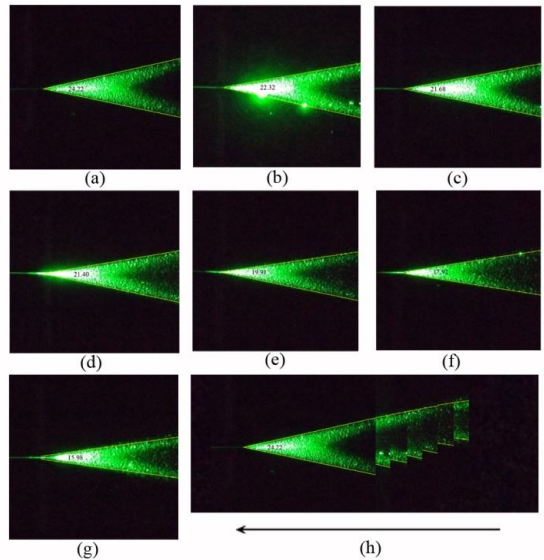

Fig. 8. (a-g) 532nm leaky radiation pattern in seven different PDMS-toluene solutions with volume ratio 1:8 , 1:6 , 1:5, 1:4 , 1:3 , 1:2.5 , 1:2. (h) Comparison of the different radiation angles for different solutions

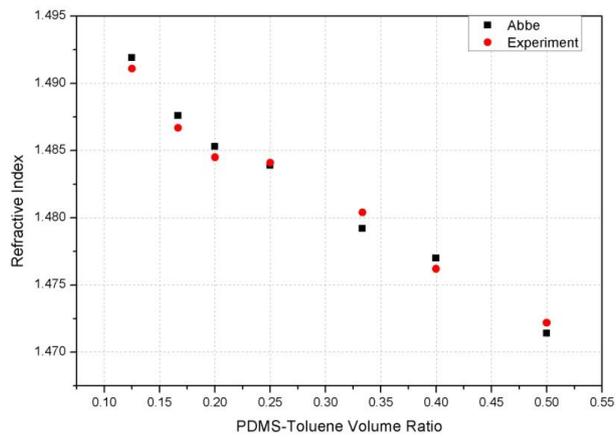

Fig. 9. Measured RI at 532nm wavelength by leaky radiation (red circle dots) and Abbe refractometer (black square dots) for different PDMS-toluene volume radio solutions.

## 5. Conclusion

A compact and broadband microfiber sensor has been realized based on its leaky radiation. The sensor shows good performance in the RI detection at different wavelength. It's a very simple and portable technique. These advantages give the sensor good opportunity for practical applications.

**Acknowledgments**

This work was financially supported by the National Natural Science Foundation of China (No. 11321063 and 11374151), the National Key Projects for Basic Researches of China (No. 2012CB933501, 2012CB921500, and 2010CB630703), the Doctoral Program of Higher Education (20120091140005), Research Grants Council Earmarked Research Grants (M-HKUST601/12), the Program for New Century Excellent Talents in University (NCET-10-0480) and Dengfeng Project B of Nanjing University.